\documentclass[aps,prl,twocolumn,english]{revtex4}
\usepackage{amssymb}
\usepackage{amsmath}
\usepackage{amsthm}
\usepackage{color}
\usepackage{graphicx}%
\makeatletter
\usepackage{babel}
\makeatother
\begin{document}
\title{Current-induced spin polarization in nonmagnetic semiconductor junctions}
\author{Yunong Qi$^1$ and Michael E. Flatt\'e$^{1,2}$}
\email{michael_flatte@mailaps.org}
\affiliation{$^1$Optical Science and Technology Center and Department of Physics
\& Astronomy, University of Iowa, Iowa City, Iowa 52242\\ $^2$ Kavli Institute for Theoretical Physics, University of California, Santa Barbara, California 93106}
\date{\today}
\begin{abstract}
Spontaneous spin polarization of the electrical current flowing through nonmagnetic semiconductor junctions can be generated by carrier scattering processes that are independent of the carrier spin. The two required elements for current-induced spin polarization are (1) the presence of built-in spatially-varying electric fields in the junction and (2) energy-dependent carrier scattering processes. Spin-orbit interactions are not required for this effect, thus it should occur in materials like silicon that lack significant spin-orbit interactions.  Approximate analytic expressions as well as detailed numerical simulations of the time-dependent nonlinear spin transport in a GaAs junction strongly suggest that the recent experimental observation of current-induced spin polarization in this system [Y. Kato, R. C. Myers, A. C. Gossard, and D. D. Awschalom, Phys. Rev. Lett. {\bf 93}, 176601 (2004)] may be explained by this effect.
\end{abstract}
\maketitle
The goal of generating a large nonequilibrium spin polarization in a nonmagnetic semiconductor has led to dramatic recent advancements in understanding the fundamentals of transport in magnetic and nonmagnetic semiconductors, and is an essential requirement for semiconductor spintronic applications.\cite{Spintronics-book,Sawolf} Efforts to achieve high-efficiency spin injection from a magnetic material into a nonmagnetic semiconductor\cite{Fiederling,Ohno,Zhu,Jonker,Crowell,Salis} have yielded $>90$\% spin polarization from magnetic semiconductor contacts and $>70$\% spin polarization from ferromagnetic metal contacts at room temperature. The recent experimental demonstration\cite{AwschalomSHE,SihNP,Wunderlich} of the spin Hall effect\cite{Dyakonov,Levitov,Edelstein,Zhang,Sinova} showed that spin polarization, both in density and current, can be induced in a nonmagnetic semiconductor through the spin-orbit interaction (without recourse to any magnetic materials). The same spin-orbit interaction that generates these currents, however, also leads to rapid spin decoherence\cite{Meier}, so the spontaneous generation of spin polarization in a material with a very long spin coherence time remains elusive. One pathway towards generating spin polarization in materials with weak spin-orbit interaction, a ``spin Gunn effect'', was suggested recently\cite{Spingunn}, but this pathway requires the semiconductor to be undergoing the rapid oscillations and moving inhomogeneous domains characterizing electrical current flow in the Gunn effect. In addition to the significant heating and microwave radiation characterizing the Gunn effect, only a very few semiconductor materials have exhibited this effect to date, and they all have strong spin-orbit interactions.

Here we identify a much broader class of nonmagnetic materials and structures for which spontaneous spin polarization will arise dynamically during carrier transport, but without the spontaneously-formed space-charge domain of the Gunn effect. The essential requirement is the presence in the semiconductor of a region of {\it static inhomogeneous electric field}, such as arises naturally when the doping of a semiconductor region is varied spatially.  The resulting spontaneous spin polarization is unrelated to the spin Hall effect, as it can arise in the complete absence of spin-orbit interactions. Thus spontaneous spin polarization should be achievable in materials with long conduction electron spin coherence times, such as is known for ZnO\cite{ghoshZnO}, and is thought to be the case for silicon.  Due to the carrier energy dependence of scattering processes, ubiquitous in ordinary semiconductors, a small excess spin polarization (thermally nucleated or seeded with a magnetic field) can be dramatically amplified during transport across the region of static electric field. This is demonstrated through the analytic and numerical analysis of the nonlinear equations for time-dependent current transport across a junction between two regions of a semiconductor with different $n$-type doping levels (an $n^+-n$ junction).  This effect may explain a very large, unexplained, experimentally-observed current-induced spin polarization in GaAs.\cite{Kato2004}
   
\begin{figure}[h]
\includegraphics[width=8cm]{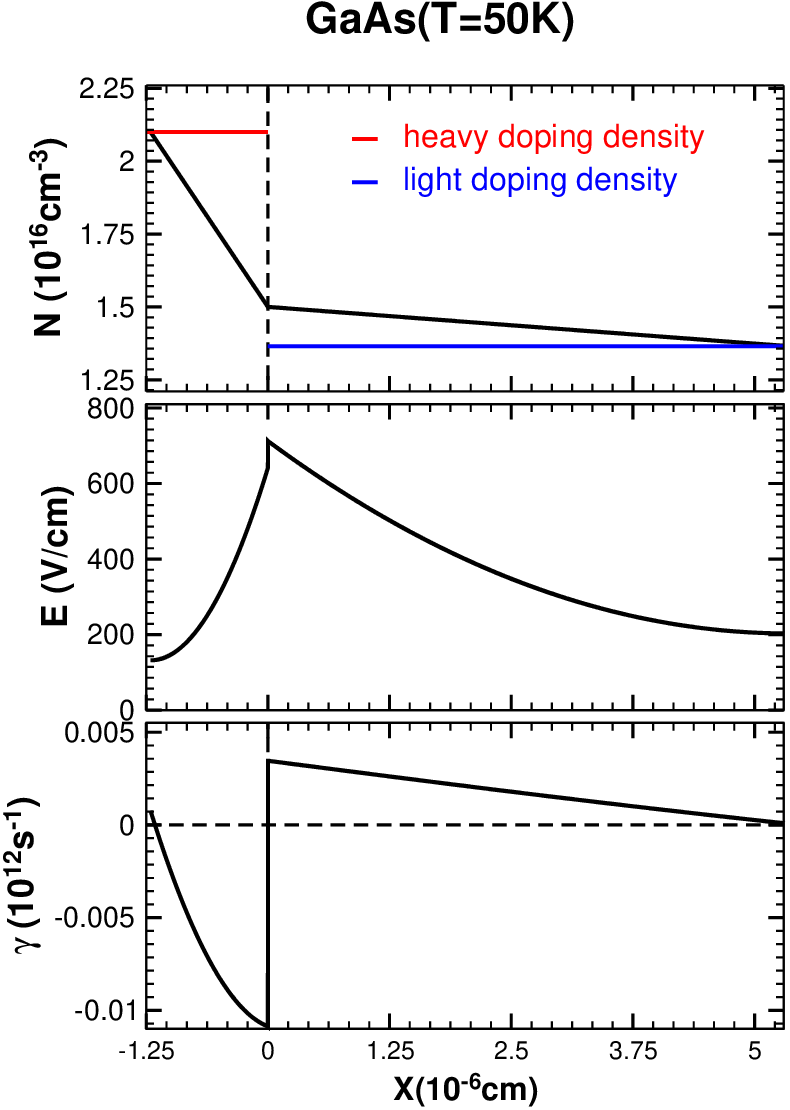}
\caption{(color online) Dependence on position of (a) carrier density, (b) electric field, and (c) spin polarization amplification rate (SPAR) for unpolarized carriers in an $n^+$--$n$ junction in GaAs at $T=50$K.}\label{schematic}
\end{figure}

The phenomenon of spin amplification of current transport through an $n^+$--$n$ junction appears to be generic, and occurs for a wide range of temperatures and doping. Although calculations here are presented for GaAs, similar calculations for other materials indicate similar spin amplification (leading to larger current-induced spin polarization in materials with smaller spin-orbit interactions). Approximate analytic expressions for the spin amplification have been found which permit a qualitative understanding of the temperature and doping dependence of the effect. Also presented here will be detailed numerical calculations for a specific GaAs junction at $50$K with a doping level of $n_{D1}=2.25{\times}10^{16}$cm$^{-3}$ for $x < 0$ and of $n_{D2}=1.35{\times}10^{16}$cm$^{-3}$ for $x>0$. Similar numerical results have been obtained for many other junctions as well. To help the numerical stability of the calculations the two doping levels considered in a junction differ by no more than a factor of two, although the analytic results indicate larger doping variations will be even more effective in generating spontaneous spin polarization.  

The larger density of carriers on the $n^+$ side of the junction leads to carrier diffusion across the interface and the growth of an inhomogeneous space-charge dipole electric field until the diffusion current is balanced by a drift current from the dipole field. This space-charge field will be the inhomogeneous electric field that drives current-induced spin polarization. The equilibrium charge carrier distribution and electrochemical potential near the interface can be determined precisely from a self-consistent Poisson equation, however  the calculations require these quantities far from equilibrium where its joint solution with the nonlinear drift-diffusion equations would be required. As we must solve dynamic nonlinear drift-diffusion equations for spin densities, already a computationally challenging task, we model the charge density through the junction with a simple analytic model\cite{Smith,Gunn-junction} as shown in Fig.~\ref{schematic}(a). The charge density is assumed to vary linearly on both sides of the interface. The two slopes and the intersection point are found by requiring global charge neutrality, continuity of the charge density across the interface, and the correct values of the electric fields at $\pm \infty$. This produces a parabolic analytic expression for the electric field near the interface. Figure~\ref{schematic}(b) shows the calculated electric field for an asymptotic field of $200$V$/$cm as $x\rightarrow \infty$.

The origin of current-induced spontaneous spin polarization
in non-magnetic semiconductors {\it without the spin-orbit interaction} is a dependence of the carrier mobility, $\mu_\uparrow$ and $\mu_\downarrow$ on the carrier spin polarization $P=(n_{\uparrow} - n_{\downarrow})/(n_{\uparrow} + n_{\downarrow})$.\cite{Spingunn} We approximate this spin polarization dependence identically to Ref.~\onlinecite{Spingunn}: 
\begin{equation}
{\mu}_{\uparrow(\downarrow)} = {\mu}(n/2)\cdot\left(1 +(-) {\alpha}P\right),\label{alphaintro}
\end{equation}
where the coefficient
$\alpha$ depends on the dominant spin-conserving carrier scattering
mechanism.  A discussion of the dependence of $\alpha$ on
carrier density, temperature, and spin-conserving scattering process can be found in Ref.~\onlinecite{Spingunn}. $\alpha$ also depends on $P$, which will cause our dynamical equations for spin polarization to become nonlinear.

To explore the consequences of the spin-polarization-dependent mobility in a static inhomogeneous electric field requires the numerical solution of spin-dependent drift-diffusion equations\cite{Spintronics-book}:
\begin{eqnarray}
\frac{{\partial}n_{{\uparrow}}}{{\partial}t}&=& {-}
\frac{n_{\uparrow} -
n_{\downarrow}}{2T_{1}} -
\frac{{\partial}\left(n_{{\uparrow}}v_{{\uparrow}}\right)}{{\partial}x} +
D_{{\uparrow}}\frac{{\partial}^{2}n_{{\uparrow}}}{{\partial}x^{2}},\\
\frac{{\partial}n_{{\downarrow}}}{{\partial}t}&=& {+}
\frac{n_{\uparrow} -
n_{\downarrow}}{2T_{1}} -
\frac{{\partial}\left(n_{{\downarrow}}v_{{\downarrow}}\right)}{{\partial}x} +
D_{{\downarrow}}\frac{{\partial}^{2}n_{{\downarrow}}}{{\partial}x^{2}},
\end{eqnarray}
where $T_{1}$ is the spin relaxation time of the electrons and $D_{\uparrow(\downarrow)}$
is the diffusion constant of the electrons for spin up(down). Substituting $\mu_{\uparrow(\downarrow)}$ into these from
Eq.~(\ref{alphaintro}), keeping only terms up to first order in $P$, and assuming $D_\uparrow=D_\downarrow$ yields
\begin{eqnarray}
\frac{{\partial}P}{{\partial}t} & = & \left[{\gamma} - \frac{1}{T_{1}}\right]P +
D\frac{{\partial}^{2}P}{{\partial}x^{2}} + \nonumber\\
&&\qquad\qquad\qquad\left[{\mu}E\left(1 -
{\alpha}\right) +
\frac{D}{n}\frac{{\partial}n}{{\partial}x}\right]\frac{{\partial}P}{{\partial}x},\label{dde-p}
\end{eqnarray}
where 
\begin{equation}
{\gamma} = -
\frac{\alpha}{n}\frac{{\partial}[{n\mu E}]}{{\partial}x}\label{gamma} \end{equation}
is the spin polarization amplification rate (SPAR) at a given position. In Ref.~\onlinecite{Spingunn} by shifting to the moving frame of the Gunn domain and neglecting carrier diffusion Eq.~(\ref{dde-p}) was reduced to $d P/dt = (\gamma-T_1^{-1})P$, which was simple to solve even for a $P$-dependent $\gamma$.

For an $n^+-n$ junction all terms in Eq.~(\ref{dde-p}) must be retained, which makes the calculation considerably more difficult, but the results much more general.  For the carrier density and electric field of the junction shown in Fig.~\ref{schematic}(a,b) the SPAR for $P=0$ is shown in Fig.~\ref{schematic}(c). For this junction the characteristic nature of the SPAR is a short region of negative values on the $n^+$ side of the interface followed by an extended region of positive values on the $n$ side of the interface. The functional form of the SPAR can be approximated as
\begin{equation}
\gamma(x)= \gamma_{\rm max} (1-x/x_a),\label{SPAR}
\end{equation}
where the length of the accumulation region (50~nm in Fig.~\ref{schematic}) is $x_a$.
The polarization of a pulse of partially polarized spins should be amplified when it travels through a region of positive SPAR (exceeding the spin relaxation rate). The maximum SPAR in Fig.~\ref{schematic}(c) is 3~GHz, which leads to amplification if the spin lifetime exceeds 300~ps. At 50K in the GaAs-based experiments of Ref.~\onlinecite{Kato2004} the spin lifetime is always at least an order of magnitude longer.

For junctions between two regions of differing doping levels, and for the more common case\cite{Spingunn} of $\alpha>0$, each of $n$, $\mu$, and $E$ are decreasing in the depletion region (see Fig.~\ref{schematic} for the $n$ and $E$ dependence; for $\alpha>0$, $\mu$ follows the $n$ dependence). With these trends (from Eq.~(\ref{gamma})) the SPAR's are positive in the depletion region. For cases with $\alpha<0$ a more detailed calculation is required to know where the SPAR's will be positive. For ionized impurity and optical phonon scattering, the common mechanisms of mobility in semiconductors at low and high temperature, $\alpha>0$.  The position dependence of the SPAR is similar for all such junctions, an initial suppression region followed by an extended region of positive values, linearly diminishing up to the edge of the accumulation region. 

\begin{figure}[h]
\includegraphics[width=8cm]{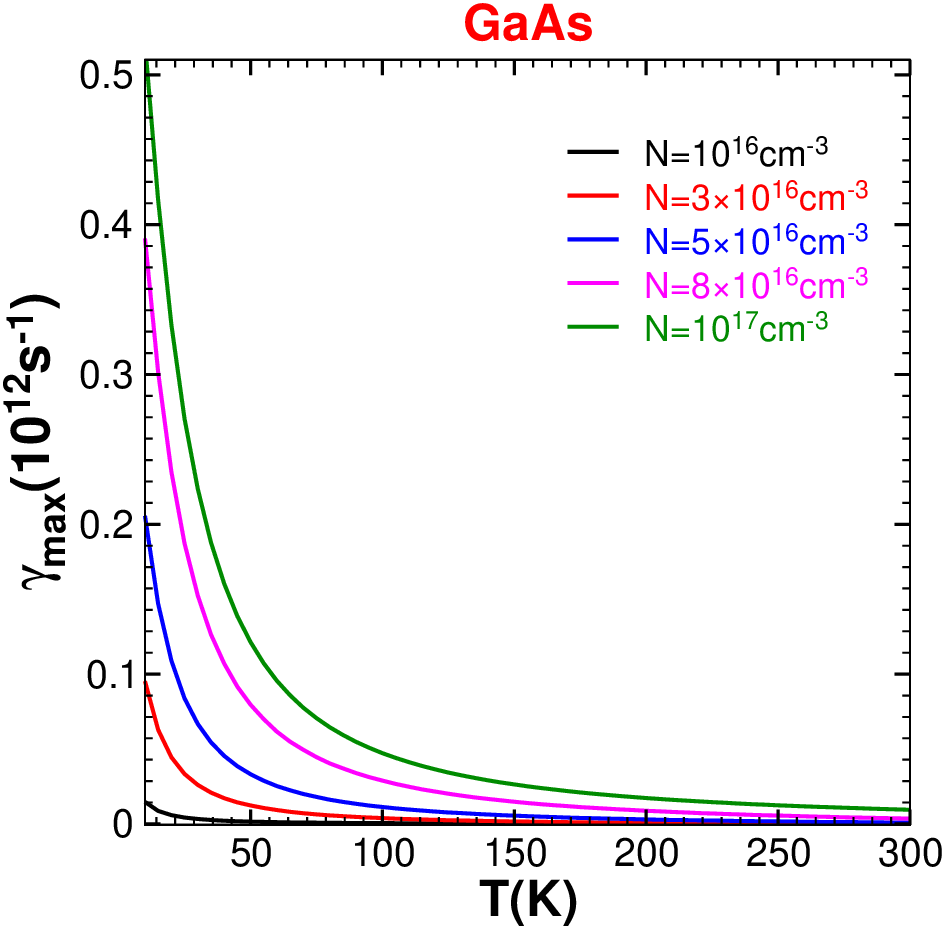}
\includegraphics[width=8cm]{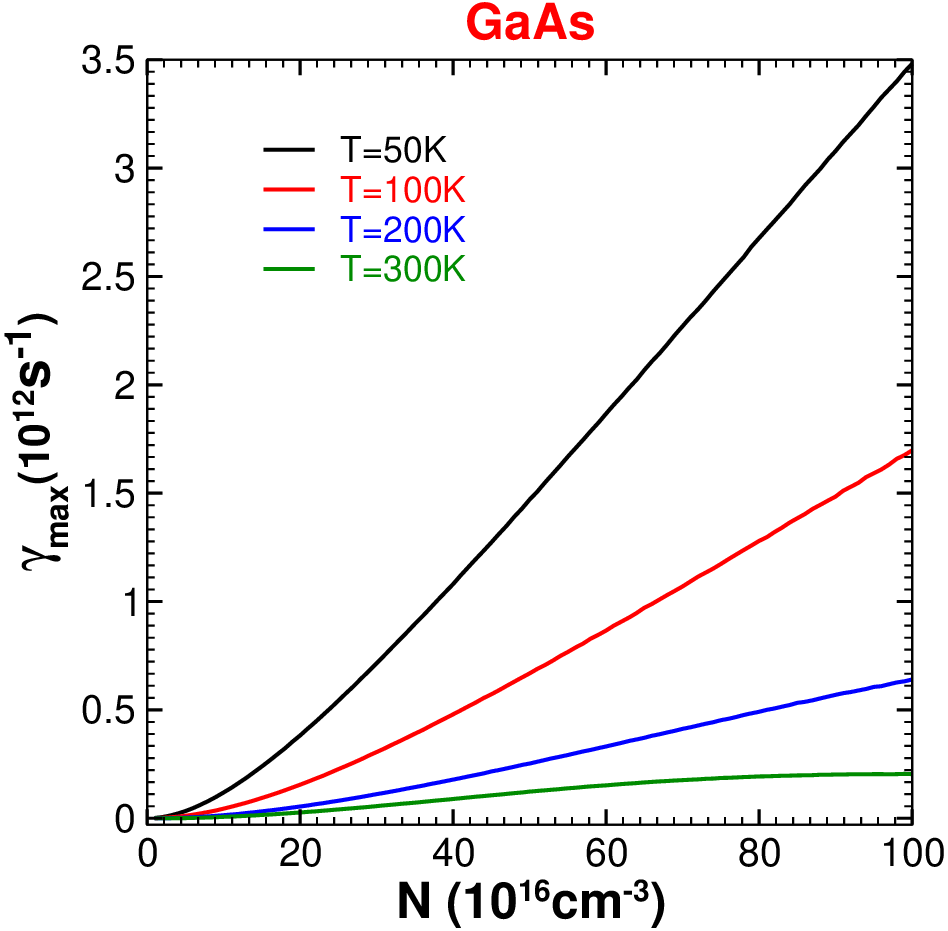}
\caption{(color online) The maximum spin-polarization amplification rate (SPAR) $\gamma_{\rm max}$, for unpolarized  carriers as a function of temperature and junction density. Ionized impurity scattering dominates the mobility.}\label{tempdep}
\end{figure}

The behavior of a wide variety of junctions can be estimated from the maximum value of the SPAR, which is shown in Fig.~\ref{tempdep}. All the $n^+-n$ junctions considered have $n_{D2}= 0.6n_{D1}$ and the mobility is considered to originate from ionized impurity scattering. As the temperature is increased the dependence of the mobility on spin diminishes ($\alpha$ decreases\cite{Spingunn}) so the SPAR diminishes as well (Fig.~\ref{tempdep}(a)). As the carrier density is increased the dependence of the mobility on spin increases ($\alpha$ increases\cite{Spingunn}) so the SPAR increases (Fig.~\ref{tempdep}(b)). At room temperature in a system doped $n=10^{18}$cm$^{-3}$ the SPAR exceeds the spin relaxation rate if $T_1>5 ps$.

For the polarization of a pulse of carriers to be significantly amplified in the junction the pulse must remain in the region of positive SPAR for sufficient time. The instantaneous velocity of a piece of a pulse of carriers, from the coefficient of the $dP/dx$ term of Eq.~(\ref{dde-p}) is 
\begin{equation}
v(x)={\mu}E\left(1 -
{\alpha}\right) + \frac{D}{n}\frac{{\partial}n}{{\partial}x}.
\end{equation}
For the junction considered in Fig.~\ref{schematic} $v(x)<\mu E$ in the region of amplification, $0<x<50$~nm, both because $\alpha>0$ and $\partial n/\partial x<0$. This self-consistent reduction in the effective velocity of a carrier pulse plays an important role in amplification, as otherwise the pulse would only remain $\sim 25$~ps in the amplification region.  An analytic expression can be derived with the assumptions of Fig.~\ref{schematic} and (1) the SPAR has the form of Eq.~(\ref{SPAR}), (2)   the polarization dependence of $\alpha$ (and thus of $\gamma$) can be neglected, and (3) diffusion (the $\partial^2P/\partial x^2$ term in Eq.~(\ref{dde-p}) can be neglected. Under these conditions $v(x)=v_{\rm eff}$ is a constant in the positive SPAR region. The polarization of the pulse would be
\begin{equation}
P(t) = P_{\rm initial} {\exp}\left[{(\gamma_{\rm max}-T_1^{-1})(t -v_{\rm eff}t^2/2x_a)}\right].
\end{equation}
Evaluating at $t=x_a/v_{\rm eff}$ yields an amplification of
\begin{equation}
\frac{P_{\rm final}}{P_{\rm initial}} = {\exp}\left[{\frac{(\gamma_{\rm max}-T_1^{-1})x_a}{2v_{\rm eff}}}\right].
\end{equation}

\begin{figure}[h]
\includegraphics[width=8cm]{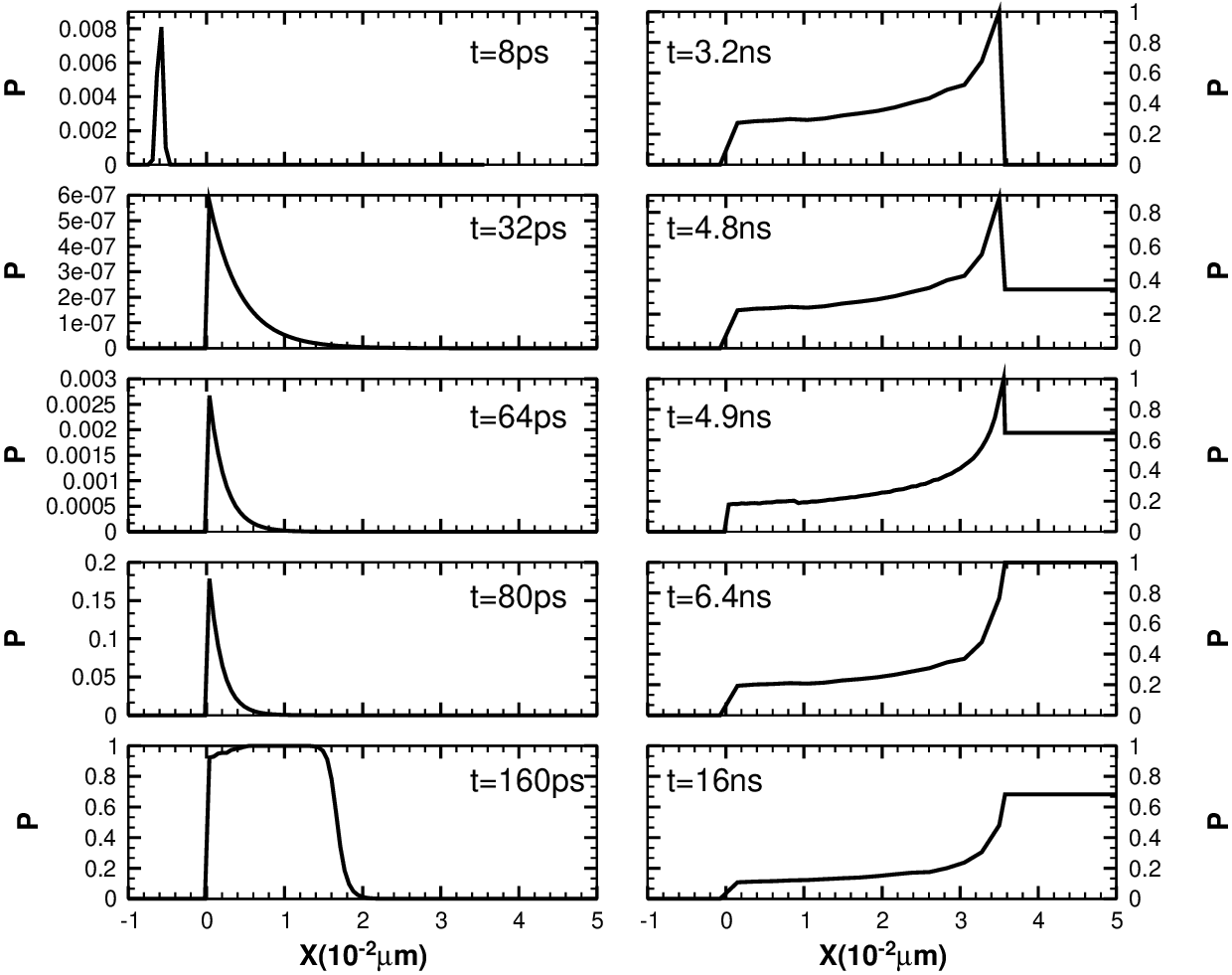}
\caption{Propagation of an initial weakly spin-polarized pulse through the $n^+-n$ junction of Fig.~1. Amplification occurs for $x>0$ and yields a nearly 100\% spin-polarized pulse into the $n$-doped region. The spatial dependence of the spin polarization is shown for 10 time slices from $8$~ps to $16$~ns.}\label{pulse}
\end{figure}

A numerical calculation that includes the spin polarization dependence of the SPAR and $v_{\rm eff}$, and also includes diffusion, for the junction in Fig.~\ref{schematic}, is shown in Fig.~\ref{pulse} at $T=50K$.
The system is initialized at $t=0$ with a weakly spin polarized
gaussian pulse centered at $x=-d_{1}$ on the $n^+$ side of the $n^{+}-n$ junction. Subsequent panels in Fig.~\ref{pulse} are labeled by the time and indicate that the pulse is suppressed as it propagates through the depletion region ($x<0$), as expected from the negative SPAR in Fig.~\ref{schematic}(c) in this region. During this the pulse drifts towards the interface and spreads due to diffusion and the position-dependent velocity $v(x)$. At $t=32$ps the spin polarization reaches the amplification region at $x>0$ and begins to increase. By  $t=160$ps the spin polarization has saturated at $P\sim 1$.  The distorted pulse then slowly drifts towards the edge of the accumulation region. Outside of the accumulation region the velocity is much larger than within, so it takes $\sim 5$ns for the highly spin polarized pulse within the accumulation region to drive a highly spin-polarized pulse out of the junction.  Then this
spin polarization pulse packet  moves away from the junction.
The output appears to be a nearly 100\% spin-polarized pulse that is several ns wide (corresponding to $\sim 10\mu$m in spatial width).

We found the spontaneous generation of highly spin-polarized pulses of carriers occurs in current flow through an $n^+-n$ junction. An analytic expression was derived for such junctions, neglecting diffusion and spin polarization saturation. Numerical results also support these findings.  Calculations showing the presence of highly spin-polarized current in GaAs junctions indicate these results provide a possible explanation for the current-induced spin polarization of Ref.~\onlinecite{Kato2004}.

This work was supported by DARPA/ARO DAAD19-01-0490.

\bibliography{}

\begin{references}
\bibitem{Spintronics-book} {\it Semiconductor Spintronics and Quantum
Computation}, edited by D. D. Awschalom, N. Samarth, and D. Loss  (Springer
Verlag, Berlin, 2002).
\bibitem{Sawolf}S. A. Wolf, D. D. Awschalom, R. A. Buhrman, J. M. Daughton, S. von
Molnar, M. L. Roukes, A. Y. Chtchelkanova, D. M. Treger, Science {\bf 294},
1488(2001).
\bibitem{Fiederling} R. Fiederling {\it et al.}, 
 Nature (London) {\bf 402}, 787 (1999);
R. Fiederling, P. Grabs, W. Ossau, G. Schmidt, and L. W.
Molenkamp, Appl. Phys. Lett. {\bf 82}, 2160(2003).
\bibitem{Ohno} Y. Ohno {\it et al.},
Nature (London) {\bf 402}, 790 (1999);
D. K. Young, E. Johnston-Halperin, D. D. Awschalom,Y. Ohno, H. Ohno,
Appl. Phys. Lett. {\bf 80}, 1598(2002); D. K. Young, J. A. Gupta,
E. Johnston-Halperin, E. Epstein, Y. Kato, and D. D. Awschalom, Semicond.
Sci. Technol. {\bf 17}, 275(2002).
\bibitem{Zhu} H. J. Zhu {\it et al.}, 
 Phys. Rev. Lett. {\bf 87}, 016601 (2001).
\bibitem{Jonker} A. T. Hanbicki, {\it et al.}, 
 Appl. Phys. Lett. {\bf 80}, 1240 (2002).
\bibitem{Crowell} A. F. Isakovic {\it et al.}, Phys. Rev. B {\bf 64}, 161304(R) (2001).
\bibitem{Salis} G. Salis {\it et al.}, Appl. Phys. Lett. {\bf 87}, 262503 (2005).
  \bibitem{AwschalomSHE} Y. K. Kato, R. C. Myers, A. C. Gossard, and D. D. Awschalom, 
 Science {\bf 306}, 1910 (2004).
 \bibitem{SihNP} V. Sih {\it et al.}, Nature Physics {\bf 1}, 31 (2005).
 \bibitem{Wunderlich} J. Wunderlich, B. Kaestner, J. Sinova, and T. Jungwirth, Phys. Rev. Lett. {\bf 94}, 047204 (2005).
\bibitem{Dyakonov} M. I. D'yakonov and V. I. Perel', 
 Phys. Lett. A {\bf 35}, 459 (1971).
\bibitem{Levitov} L. S. Levitov, Yu. V. Nazarnov, and G. M. \'Eliashberg, 
 Zh. Eksp. Teor. Fiz. {\bf 88}, 229 (1985) [ Sov. Phys. JETP {\bf 61}, 133-135 (1985)].
\bibitem{Edelstein} V. M. Edelstein,
 Solid State Comm. {\bf 73}, 233 (1990).
\bibitem{Zhang} S. Murakami, N. Nagaosa, and S.-C. Zhang, 
 Science {\bf 301}, 1348 (2003).
\bibitem{Sinova} J. Sinova, {\it et al.}, Phys. Rev. Lett. {\bf 92}, 126603 (2004).
\bibitem{Meier}\emph{Optical Orientation}, edited by F. Meier and B. P. Zakharchenya 
(North-Holland, New York, 1984). 
\bibitem{Spingunn} Y. Qi, Z. G. Yu, M. E. Flatt\'e, Phys. Rev. Lett. {\bf 96}, 026602 (2006).
\bibitem{ghoshZnO} S. Ghosh {\it et al.}, Appl. Phys. Lett. {\bf 86}, 232507 (2005).
\bibitem{Kato2004}Y. K. Kato, R. C. Myers, A. C. Gossard, and D. D. Awschalom, Phys. Rev. Lett. {\bf 93}, 176601 (2004). 
 \bibitem{Smith} See, e.g., 
{\it Semiconductors,} R. A. Smith (Cambridge University Press, New York, 1978).
\bibitem{Gunn-junction} J. B. Gunn, J. Electron. Control. {\bf 4}, 17 (1958). 


\end{references}

\end{document}